# AI-Driven Smart Sportswear for Real-Time Fitness Monitoring Using Textile Strain Sensors

Chenyu Tang, Wentian Yi, Zibo Zhang, Edoardo Occhipinti, and Luigi G. Occhipinti, *Senior Member, IEEE*

***Abstract*—Wearable biosensors have revolutionized human performance monitoring by enabling real-time assessment of physiological and biomechanical parameters. However, existing solutions lack the ability to simultaneously capture breath-force coordination and muscle activation symmetry in a seamless and non-invasive manner, limiting their applicability in strength training and rehabilitation. This work presents a wearable smart sportswear system that integrates screen-printed graphene-based strain sensors with compact electronics for wireless data transfer and a deep learning framework for real-time classification of exercise execution quality. By leveraging 1D ResNet-18 for feature extraction, the system achieves 92.1% classification accuracy across six exercise conditions, distinguishing between breathing irregularities and asymmetric muscle exertion. Additionally, t-SNE analysis and Grad-CAM-based explainability visualization confirm that the network accurately captures biomechanically relevant features, ensuring robust interpretability. The proposed system establishes a foundation for next-generation AI-powered sportswear, with applications in fitness optimization, injury prevention, and adaptive rehabilitation training.**

## I. INTRODUCTION

Wearable biosensors have significantly advanced human health and fitness monitoring by enabling continuous, real-time tracking of physiological signals in daily life [1-5]. Strength training exercises, such as bench pressing, require precise muscle activation symmetry and breath-force coordination to optimize performance and prevent injury. Improper execution, including imbalanced force exertion or uncoordinated breathing patterns, can lead to chronic musculoskeletal issues and reduced training efficiency [6, 7]. Therefore, there is a growing demand for real-time biomechanical feedback systems to enhance training quality and mitigate injury risks.

Traditional fitness monitoring approaches rely on optical motion capture, inertial measurement units (IMUs), and electromyography (EMG) sensors [8-12]. Optical systems, though highly accurate in controlled environments, are impractical for real-world gym settings due to occlusion issues and high costs. IMU-based solutions provide kinematic data but cannot directly assess muscle activation or breathing patterns, limiting their utility in evaluating force symmetry and coordination [13, 14]. While EMG sensors offer insights into muscle activity, they require skin preparation, precise electrode placement, and stable contact, making them susceptible to motion artefacts and signal degradation during dynamic workouts [15-17]. Furthermore, commercial fitness wearables primarily focus on step counts, heart rate, heart rate variability, breathing rate and general movement patterns, failing to provide exercise-specific biomechanical feedback crucial for strength training optimization [18, 19].

Despite advancements in wearable biosensing, existing fitness monitoring systems still lack real-time, high-resolution feedback on both muscle activation and breath coordination. Addressing this limitation requires innovations in both sensor technology and computational analysis.

Strain sensors have recently emerged as promising tools for wearable biomechanical monitoring, offering a non-invasive, flexible, and scalable alternative to traditional motion capture or electromyography (EMG)-based systems. However, their real-world deployment in fitness tracking remains hindered by two major technical challenges. First, many strain sensors exhibit limited sensitivity, making it difficult to capture the subtle mechanical deformations associated with muscle contractions and breathing patterns. In strength training, precise force exertion and breath synchronization play a critical role in optimizing performance, and an insufficiently sensitive sensor may fail to detect these fine-scale biomechanical variations [20, 21]. Second, strain sensors are often prone to motion artefacts and signal crosstalk, particularly in dynamic exercises where multiple body segments move simultaneously [22, 23]. External mechanical disturbances, such as fabric displacement, unintended body motions, or external pressure from gym equipment, can introduce noise, reducing the reliability of sensor readings. This issue is particularly pronounced in wearable applications, where movement artefacts can obscure the true muscle activation signals.

Beyond sensor hardware, accurately classifying and interpreting biomechanical signals presents additional computational challenges.

Traditional rule-based or threshold-driven methods for

This work involved human subjects. Approval of the ethical, experimental procedures and protocols was granted by Department of Engineering Ethics Committee, University of Cambridge, under the reference number 394. This work was supported by UKIERI, British Council UK (Contract No. 45371261). E.O. was supported by UKRI Centre for Doctoral Training in AI for Healthcare (EP/S023283/1) *(Corresponding author: Luigi G. Occhipinti).*
Chenyu Tang, Wentian Yi, Zibo Zhang, and Luigi G. Occhipinti are with Department of Engineering, University of Cambridge, Cambridge, CB3 0FA UK. Edoardo Occhipinti is with UKRI CDT in AI for Healthcare, Imperial College London, London, SW7 2AZ. (e-mail: ct631@cam.ac.uk; wy278@cam.ac.uk; zz534@cam.ac.uk; eo816@ic.ac.uk; lgo23@cam.ac.uk).

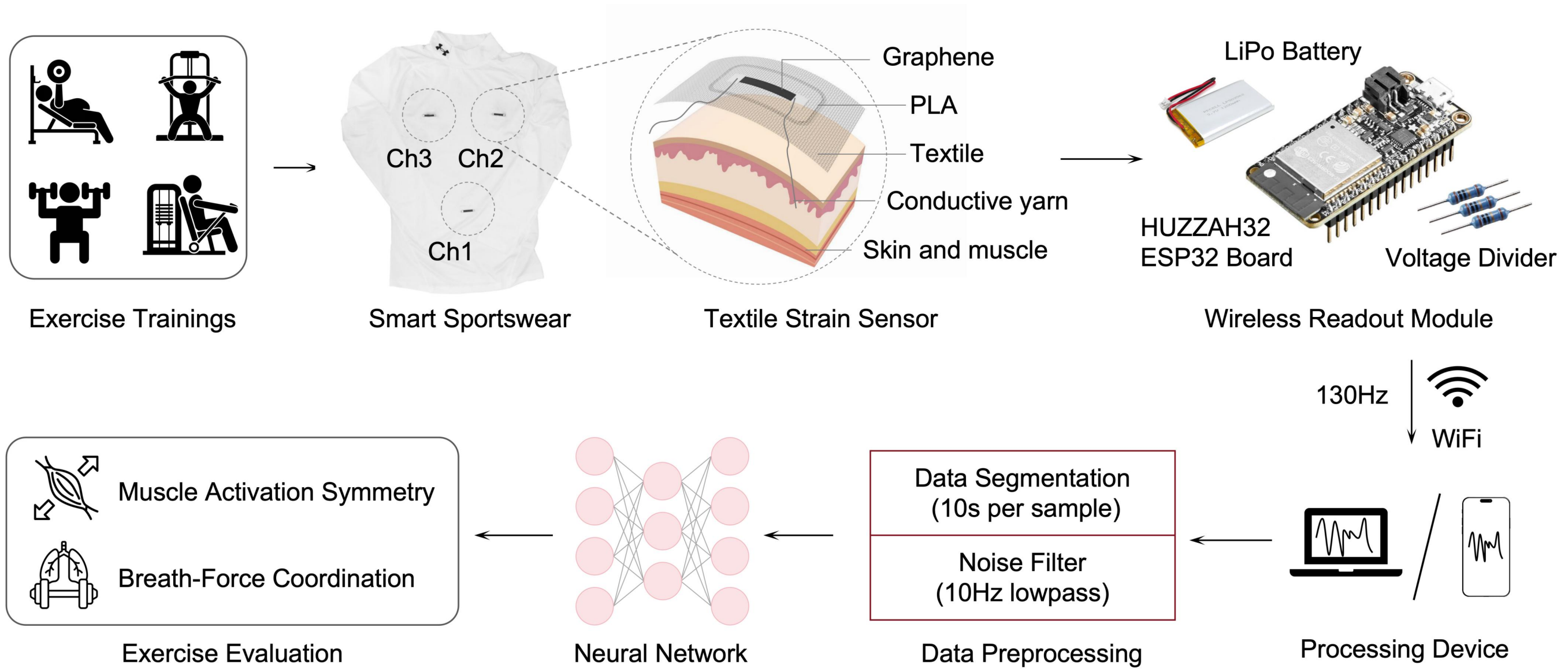

**Fig. 1. Overview of the smart sportswear system for real-time biomechanical monitoring.** The system integrates ultrasensitive textile strain sensors, screen-printed onto a compression shirt, to track breathing patterns and muscle activation symmetry during exercise. The graphene-based strain sensors detect mechanical deformations associated with breath-force coordination and upper-body muscle engagement. Signals are acquired by a wireless readout module, which includes an ESP32-based processing unit, a LiPo battery, and a voltage divider circuit for signal conditioning. The digitized signals are transmitted via Wi-Fi (130 Hz sampling rate) to a connected processing device for data segmentation, noise filtering, and neural network-based classification. The system provides real-time feedback on breathing irregularities and asymmetric force exertion, supporting improved exercise execution and injury prevention.

exercise monitoring rely on predefined criteria, such as detecting asymmetries based on fixed strain amplitude differences or breath timing thresholds. However, these methods struggle to generalize across different users due to inter-subject variability. Individual differences in muscle strength, training experience, and anatomical structure mean that a single threshold or rule set may not be applicable to all users, leading to false positives or negatives in detecting improper form. To overcome this limitation, some researchers have employed deep learning-based classification frameworks to automatically learn complex biomechanical patterns and adapt to diverse user profiles [24-26]. Despite the effectiveness of deep learning in classification, a persistent challenge in AI-driven fitness monitoring is the lack of explainability in model predictions. Neural networks often function as black-box systems, where decisions are made based on complex feature mappings that are not directly interpretable by users. This lack of transparency poses a barrier to adoption, as athletes and trainers may be hesitant to trust or act upon recommendations without a clear understanding of how the system arrives at its conclusions. To address this, we integrate Grad-CAM-based Explainable AI (XAI) techniques, allowing users to visualize the specific signal features that influenced classification outcomes [27-29].

To address these limitations, we propose a smart sportswear system that integrates screen-printed strain sensors into commercial compression garments for real-time monitoring of breathing patterns and muscle activation symmetry (see detailed diagram in Figure 1). This system enables seamless biomechanical assessment during strength training, providing instantaneous feedback on exercise execution quality without the need for external motion capture or cumbersome wearable devices. The sensor layout consists of multi-channel strain sensors printed on the abdominal and upper chest regions, allowing for simultaneous monitoring of breath coordination and left-right muscle engagement. The abdominal strain sensor captures inhalation and exhalation cycles, offering insights into breath-force synchronization during lifting, while the pectoral strain sensors detect asymmetries in muscle force application, identifying unbalanced exertion that could lead to training inefficiencies or injuries.

A key element of this system lies in its graphene-based strain sensors, which leverage an ordered micro-crack structure to achieve highly tunable sensitivity, which has been proved in our previous research [30]. This design enables the sensors to detect small-scale mechanical deformations corresponding to breathing and muscle contractions, ensuring a high signal-to-noise ratio even in dynamic exercise environments. Additionally, the system incorporates a PLA (polylactic acid) isolation layer, which minimizes motion artefacts and external mechanical disturbances, thereby improving the accuracy and reliability of captured biomechanical data. The effectiveness of this technique has also been proved in our previous study [31]. These advancements in sensor design and material engineering allow for high-resolution, artefact-resistant physiological monitoring, making the system suitable for both controlled laboratory studies and real-world smart sportware applications for use in gym exercises.

To process and analyze the acquired signals, we employ a neural network-based classification framework, which distinguishes between correct and improper exercise execution. The model is trained on data from five subjects performing both standard and five types of improper bench press movements, covering common biomechanical faults such as asymmetric muscle activation and uncoordinated breathing. Unlike traditional rule-based or threshold-driven classification methods, our deep learning approach autonomously learns complex biomechanical patterns from the strain sensor signals, enabling robust and adaptive assessment across different individuals. Experimental results demonstrate that the proposed system achieves an average classification accuracy of 92.1%, outperforming conventional heuristic-based fitness monitoring techniques.

Beyond classification accuracy, we also address the lack of explainability in deep learning-based exercise analysis by integrating Grad-CAM-based Explainable AI (XAI) techniques. This method generates visual saliency maps that highlight the key signal features contributing to the classification decision, providing users and trainers with an interpretable understanding of movement quality. By revealing which aspects of the strain sensor data influence the network's decision-making, this explainability framework enhances user trust and practical usability, bridging the gap between automated assessment and human intuition in sports training.

The proposed smart sportswear system extends beyond gym-based training, offering applications in rehabilitation, injury prevention, and daily health monitoring. Its scalable sensor fabrication allows adaptation for different muscle groups, making it suitable for stroke recovery, remote physiotherapy, and posture correction. By providing real-time, non-invasive feedback on muscle activation and breathing, it aids clinicians in tracking patient progress and helps athletes detect force asymmetries to prevent injuries. Through AI-driven biomechanical analysis and wireless real-time data acquisition, this system bridges the gap between wearable sensing and intelligent exercise guidance, paving the way for personalized, next-generation training solutions.

## II. METHODOLOGY

### A. Hardware of the System

The proposed smart sportswear system integrates ultrasensitive textile strain sensors and a wireless readout module for real-time biomechanical monitoring during exercise. The system consists of (i) screen-printed graphene-based strain sensors embedded in sportswear to detect muscle activation and breathing patterns, and (ii) a wireless ESP32-based module that enables signal acquisition, preprocessing, and Wi-Fi transmission at 130 Hz. This lightweight, untethered design ensures unrestricted movement while providing continuous feedback on exercise form and muscle symmetry.

The following sections detail the textile strain sensor integration and the wireless data acquisition module, highlighting their roles in ensuring high-sensitivity monitoring and seamless real-time feedback.

#### *1) Sportswear printed with ultrasensitive textile strain sensors*

The proposed smart sportswear integrates ultrasensitive graphene-based strain sensors into a polyester/spandex blended elastic fabric (commercially available sportswear, Under Armour, Inc.), enabling real-time monitoring of skin deformation induced by muscle activity. The sensors resort to a structured microcrack-based design, which provides exceptional sensitivity, breathability, and durability, making them well-suited for long-term wear during exercise [30, 31].

A layered structural design strategy was employed to ensure optimal sensor performance and robustness under repeated mechanical deformation (Figure 1). The strain-sensitive graphene layer was screen-printed onto the fabric at ergonomically determined locations over the pectoralis major (chest) and upper abdominal region, corresponding to key sites for monitoring muscle activation and breathing patterns. To enhance mechanical stability, a gradient modulus interface was formed by thermocompression molding at 180°C and 0.5 MPa, bonding a high-modulus PLA (polylactic acid) layer with the base fabric [32]. This interface attenuates strain artefacts introduced during initial wear, ensuring that the sensor response remains primarily driven by biomechanical activity rather than garment-induced deformation.

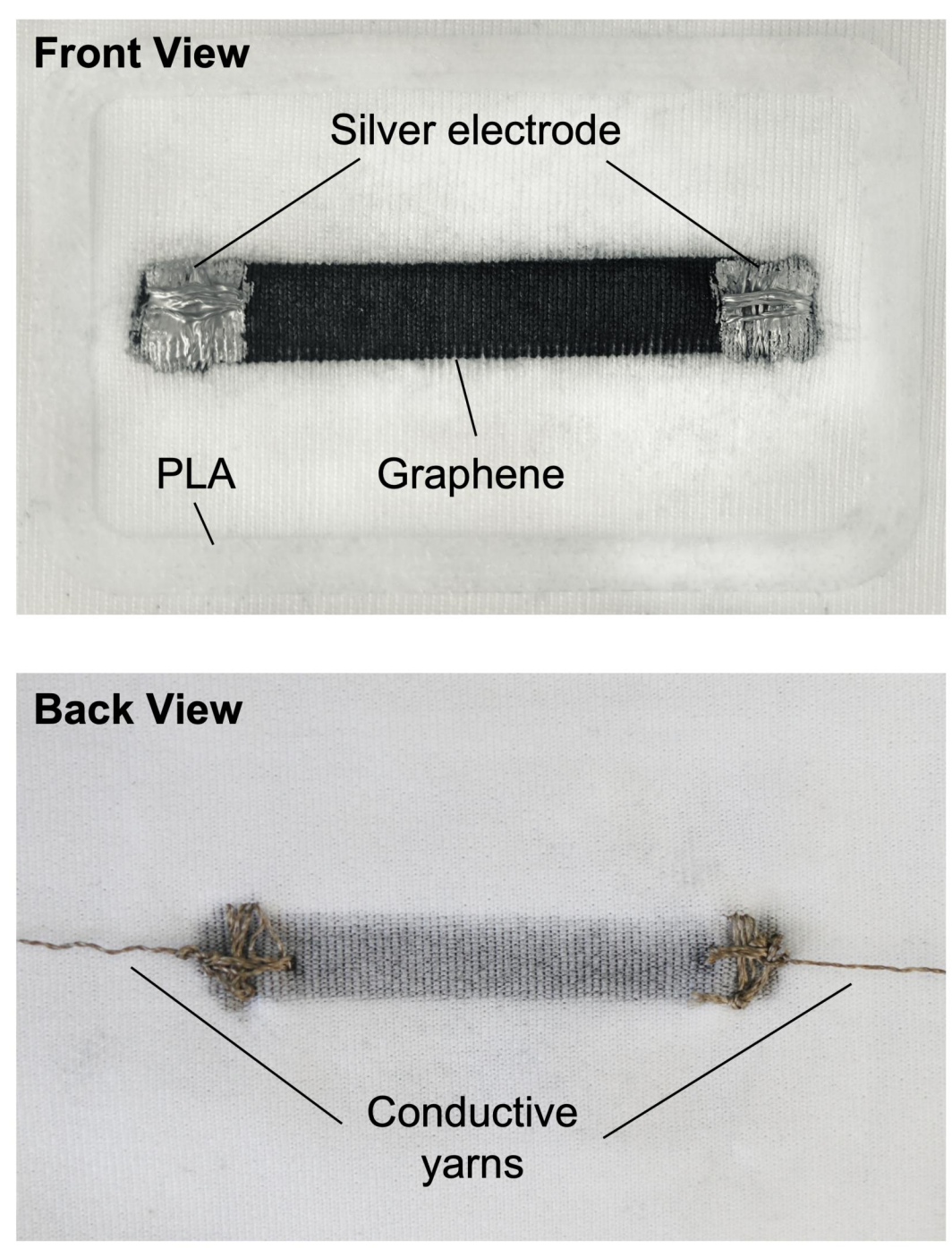

**Fig. 2.** Front and back views of a sensor channel.

To secure sweat resistance and signal stability, a hydrophobic PDMS (polydimethylsiloxane) protective layer was applied to the inner surface of the sensing region, preventing moisture-induced conductivity variations in the printed graphene coating [33]. Electrical connectivity was achieved using multi-stranded twisted silver yarn, which acted as flexible conductive wiring. These silver yarn electrodes were aligned along the fabric texture direction and secured using a wave-shaped fixation pattern, preventing additional stress accumulation that could interfere with strain measurements. Figure 2 presents a close-up photograph of a sensor channel in the smart sportswear.

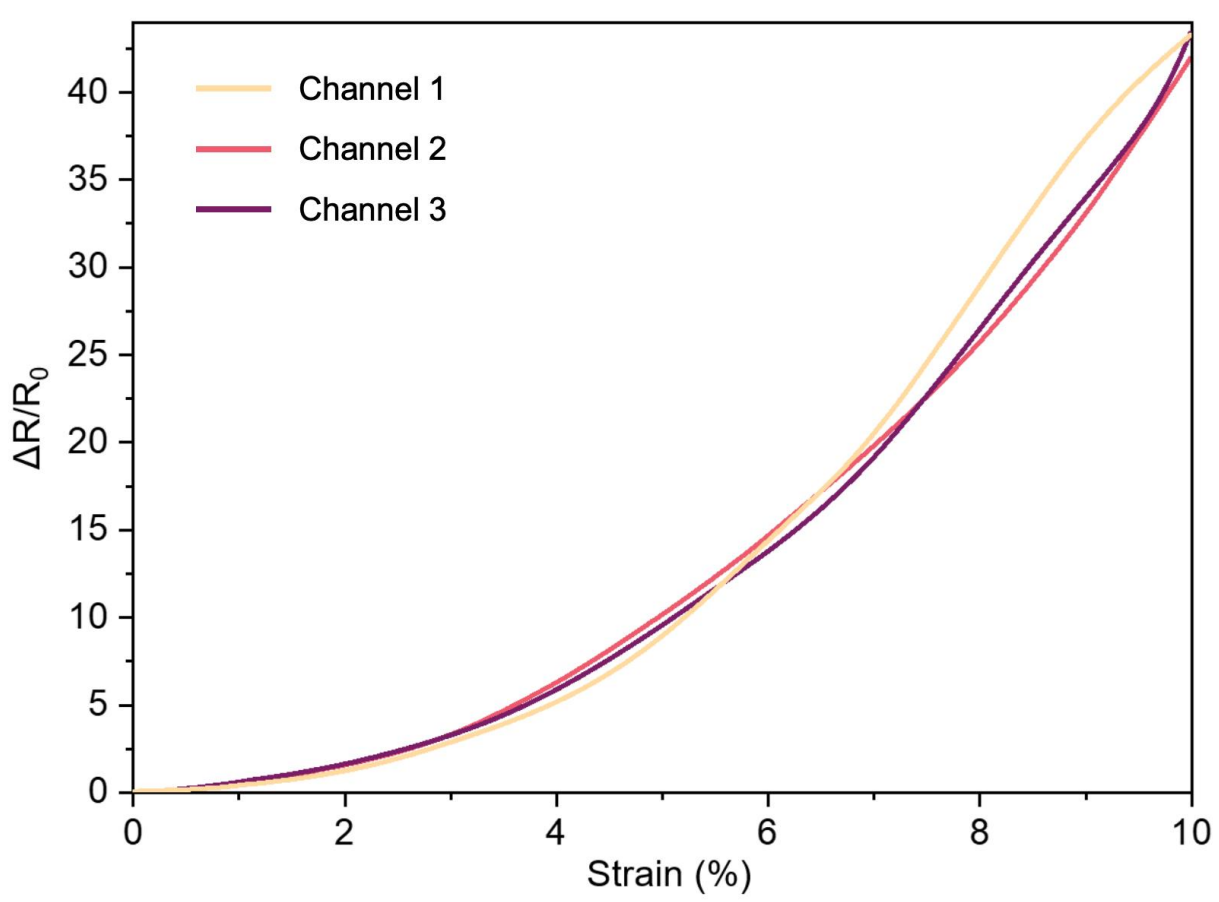

**Fig. 3. Quasi-static tensile test results for the printed graphene-based strain sensors integrated into the smart sportswear.**

To validate the mechanical response and consistency of the printed strain sensors, quasi-static uniaxial tensile tests were conducted on the three sensing channels within a 0–10% strain range. As shown in Figure 3, the sensors exhibited a highly consistent resistance-strain response across all channels, demonstrating their reliability for multi-site biomechanical monitoring. The strain sensors achieved an exceptionally high gauge factor (GF) of 433, confirming their suitability to detect minute muscle contractions and respiratory-induced skin deformations with high precision.

*2) Readout Module and Wireless Data Transmission*

To enable real-time and untethered monitoring of biomechanical signals, the proposed smart sportswear system incorporates a wireless readout module based on the Adafruit HUZZAH32 ESP32 Feather Board. This module is responsible for analog signal acquisition, preprocessing, and wireless data transmission, ensuring seamless integration with external processing devices such as laptops and smartphones. The integration of a lightweight, low-power LiPo battery allows for extended operation without restricting user movement, making the system well-suited for gym-based and free-motion exercise monitoring.

The strain sensors embedded in the textile produce resistance-based analog signals, which are first conditioned through a voltage divider circuit before being digitized by the ESP32's built-in 12-bit analog-to-digital converter (ADC). The ESP32 was selected due to its high-performance dual-core processing, low power consumption, and integrated Wi-Fi connectivity, enabling efficient real-time data streaming without requiring bulky external data acquisition systems. The module operates at a sampling rate of 130 Hz, which is sufficient to capture both muscle activation dynamics and breathing cycles with high fidelity while maintaining low latency in wireless transmission.

Wireless communication is facilitated via Wi-Fi transmission, allowing real-time streaming of processed sensor data to a connected computer or mobile device. The system architecture supports low-latency data transfer, ensuring that the feedback on muscle activation and breathing coordination can be delivered without perceptible delay. This wireless implementation eliminates the need for wired connections, providing users with greater freedom of movement and ensuring that the sportswear remains non-intrusive during exercise.

By leveraging a compact and energy-efficient readout module, combined with wireless data streaming capabilities, the proposed system achieves real-time, high-fidelity monitoring of exercise biomechanics without compromising user mobility. This enables instantaneous feedback on workout execution, facilitating real-time form correction and enhancing the applicability of smart sportswear for fitness training, rehabilitation, and remote exercise monitoring.

*B. Software of the System*

*1) Preprocessing*

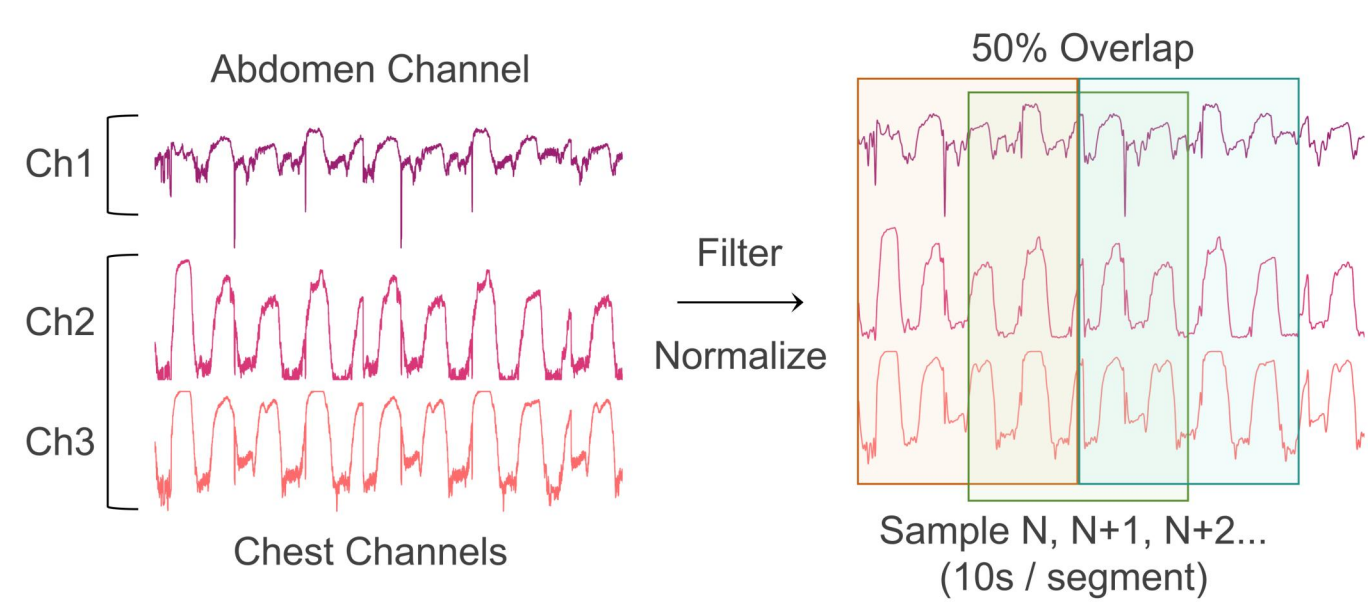

**Fig. 4. Overview of the signal preprocessing pipeline.**

To ensure high-quality input signals for neural network analysis, the collected strain sensor data undergoes a two-stage preprocessing pipeline consisting of low-pass filtering, normalization, and segmentation (see Figure 4). Given the presence of high-frequency noise introduced by motion artefacts and environmental disturbances, a finite impulse response (FIR) low-pass filter with a 10 Hz cutoff frequency is applied to remove unwanted signal fluctuations while preserving the essential biomechanical information. The filter is designed using the Hamming window function, which provides a balance between stopband attenuation and transition bandwidth, effectively suppressing high-frequency noise while minimizing distortion in the passband. The FIR filter implementation ensures zero-phase distortion by applying it in a forward-backward manner (filtfilt function), which is critical for preserving the temporal integrity of breathing and muscle activation waveforms.

Following filtering, the signals are Z-score normalized to standardize amplitude variations across different subjects and recording conditions [34]. Transforming the signals to a standard normal distribution ensures that the neural network receives consistent input signals, which helps reduce inter-subject variability and enhances model generalization ability.

After normalization, the preprocessed signals are segmented into fixed-length samples for deep learning-based classification. A sliding window approach with a 50% overlap is employed to maintain temporal continuity while increasing the number of training samples available for the model. Each window spans 10 seconds, which is sufficient to capture complete breath cycles and muscle contraction patterns. This segmentation strategy enhances classification robustness by preserving contextual information across overlapping windows, allowing the model to better differentiate between proper and improper exercise execution.

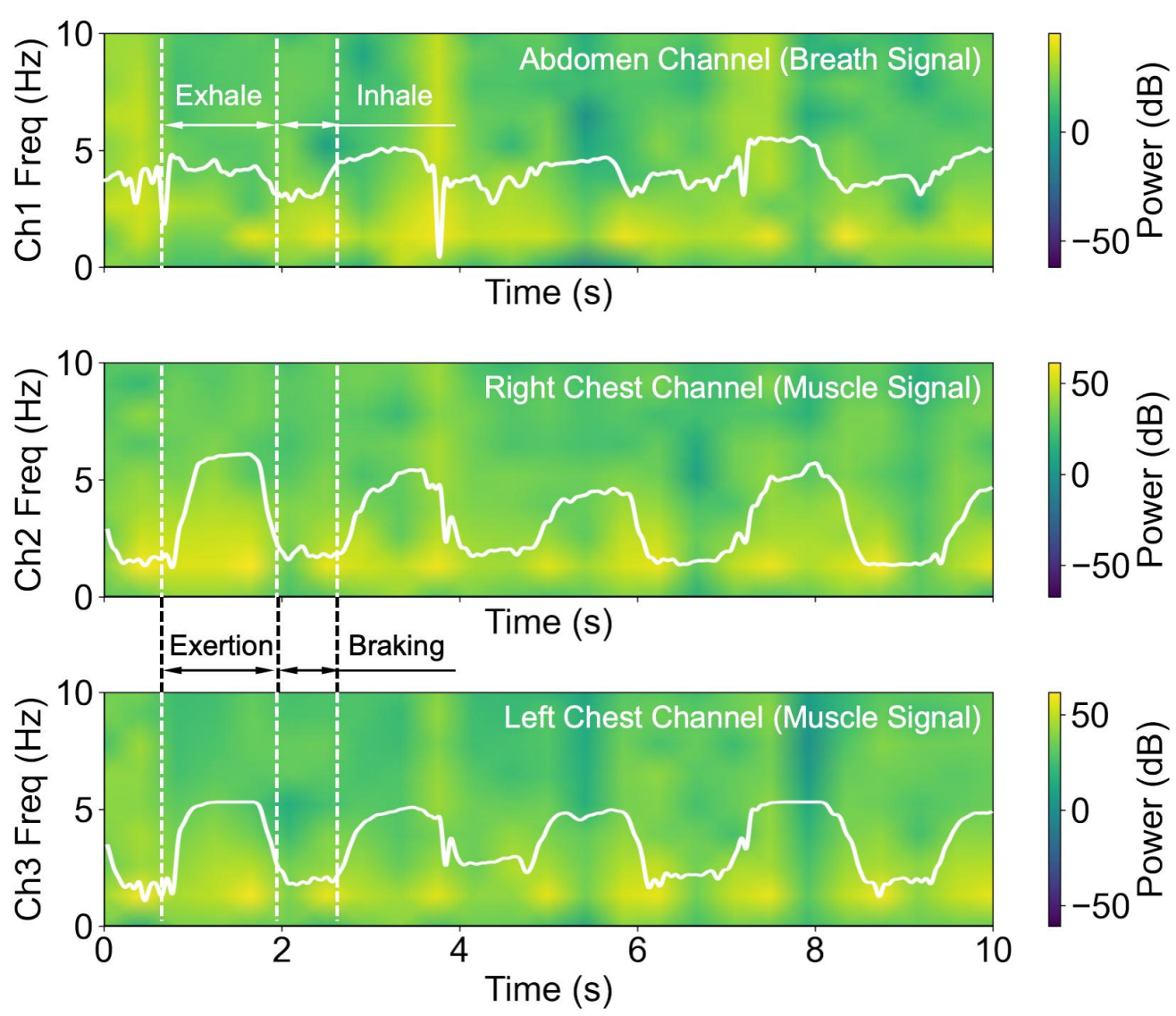

**Fig. 5. Time-frequency analysis of a properly executed bench press repetition.**

To further illustrate the characteristics of the preprocessed signals, Figure 5 presents a time-frequency visualization of a well-coordinated breath-force execution during a bench press movement. The spectrograms correspond to three sensor channels: the abdomen channel (breath signal) and the left and right chest channels (muscle activation signals).

In the abdomen channel, distinct inhale-exhale cycles are visible, with exhalation aligning with the exertion (muscle contraction) phase of lifting or pushing a weight and inhalation occurring during the braking phase, while controlling the weight release. This pattern reflects proper breath-force coordination, which is essential for maintaining stability and optimizing muscle engagement. The right and left chest channels exhibit synchronized periodic bursts of power, corresponding to pectoralis major activation during each press cycle. The exertion phase is characterized by a rise in muscle activation frequency components, which then decrease during the braking phase. The symmetry between the left and right channels indicates balanced muscle engagement, ensuring equal force distribution across both arms.

This time-frequency analysis highlights the ability of the proposed system to capture biomechanical patterns in both breathing and muscle activation. By leveraging such representations, the deep learning model can accurately differentiate between proper and improper exercise execution, improving feedback quality and exercise safety.

*2) Algorithm Development*

To classify exercise execution quality based on the acquired strain sensor signals, we develop a deep learning-based classification pipeline that leverages 1D residual networks (ResNet-18) for feature extraction and a fully connected (FC) classifier for decision-making (pipeline shown in Figure 6) [35]. The algorithm is designed to effectively capture temporal dependencies and spatial correlations within the multi-channel strain signals while maintaining robustness to inter-subject variability.

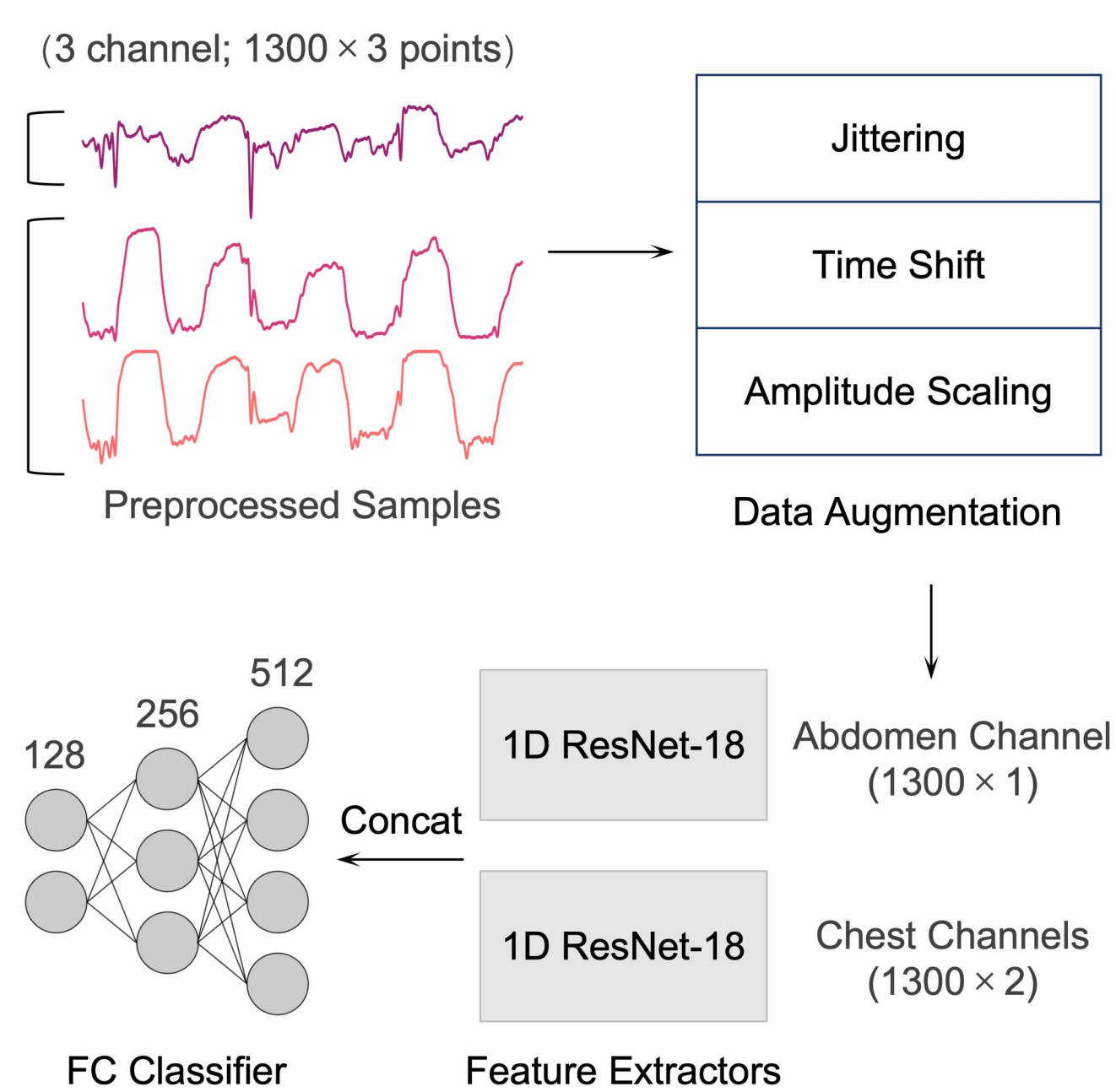

**Fig. 6. Neural network pipeline for exercise execution classification using multi-channel strain sensor signals.**

The input to the network consists of preprocessed 3-channel strain sensor signals, where each sample has a fixed length of 1300 time points (corresponding to a 10s segment at 130 Hz). To improve model generalization and reduce overfitting, we apply data augmentation techniques, including jittering (random noise injection), time shifts (phase perturbation), and amplitude scaling (variability enhancement). These augmentations help simulate real-world variations in breathing patterns and muscle activation, ensuring the model remains robust to natural deviations in exercise execution [36].

Given the distinct functional roles of the abdomen channel (breathing signal) and the chest channels (muscle activation signals), we adopt a dual-branch feature extraction strategy. The abdomen channel is processed by one 1D ResNet-18 module, while the chest channels are fed into a separate 1D ResNet-18 network. This approach allows the network to learn specialized features for breath-force coordination and muscle activation patterns, respectively. The extracted feature representations from both branches are then concatenated and passed through a three-layer fully connected classifier, which outputs the final classification decision regarding exercise execution quality.

This architecture ensures that the model can effectively distinguish between proper and improper exercise forms, including breath-holding, uncoordinated breathing, and asymmetric force exertion. The deep residual network design enhances gradient flow during training, improving convergence stability and enabling the model to learn complex temporal dependencies in the strain sensor data.

To optimize model performance, the classification pipeline was trained using a cross-entropy loss function, given the multi-class nature of the exercise classification task. The network was trained using the Adam optimizer with an initial learning rate of 0.001, which was reduced by a factor of 0.1 every 20 epochs based on validation loss convergence. The model was trained for 100 epochs with a batch size of 32, ensuring a balance between computational efficiency and gradient stability. Hyperparameters were selected through random search over a bounded space (learning rate $\in$ [1e-4, 1e-2], batch size $\in$ {16, 32, 64}, L2 $\lambda$ $\in$ [1e-5, 1e-3], followed by empirical tuning based on validation accuracy.

To evaluate generalization performance, the dataset was randomly split into 70% training, 15% validation, and 15% testing sets, ensuring that all subjects were included in both training and validation but with no overlap in test data. Early stopping with a patience of 10 epochs was applied to prevent overfitting, and L2 weight regularization ( $\lambda$ = 0.0001) was introduced to improve model robustness. Augmented data samples were used in both training and validation, while the test set consisted solely of raw, unaugmented samples to ensure a fair evaluation of real-world performance.

## III. Results

### A. Dataset Collection

To evaluate the performance of the proposed smart sportswear system, we conducted data collection with eight subjects (5 males, 3 females) who have no history of muscle disorders and varying levels of strength training experience (1-7 years). The subject demographics are summarized in Table I.

TABLE I
SUBJECT DEMOGRAPHIC DETAILS

| ID | Age | BMI ($kg/m^2$) | Max Bench Press (kg) | Years of Training | Gender |
|---|---|---|---|---|---|
| 1 | 23 | 22.9 | 60 | 3 | Male |
| 2 | 29 | 24.1 | 80 | 5 | Male |
| 3 | 21 | 23.7 | 90 | 7 | Male |
| 4 | 25 | 23.0 | 60 | 4 | Male |
| 5 | 27 | 24.0 | 70 | 6 | Male |
| 6 | 26 | 20.8 | 40 | 2 | Female |
| 7 | 25 | 22.2 | 50 | 4 | Female |
| 8 | 29 | 25.4 | 35 | 1 | Female |

Each subject performed multiple bench press repetitions while wearing the smart sportswear system in a controlled gym environment. To ensure high-quality signal acquisition, the following data exclusion criteria were established:

(1) Sensor Disconnection: If a researcher visually observed a sensor detachment or unstable signal fluctuations, the trial was excluded.

(2) Form Disruptions: Any unexpected movement interruptions or loss of controlled motion resulted in the exclusion of that specific trial.

Each bench press repetition was manually labeled by a professional strength training coach, who visually monitored the subject's execution in real time, assessing both breath-force coordination and muscle activation symmetry. The expert annotations served as ground truth labels for model training and evaluation, ensuring that the dataset accurately captured key biomechanical variations in lifting form.

Repetitions were classified into six categories based on observed breathing patterns and muscle force distribution. The proper form category included repetitions where the subject exhaled during exertion and inhaled during braking (labelled as "Even"), while maintaining symmetric left-right muscle activation without noticeable compensatory movements (labelled as "Bal", for balanced). In cases where breath-force coordination remained correct but force distribution was asymmetric, repetitions were further categorized as left-dominant or right-dominant asymmetry, depending on which side exerted greater force. Additionally, trials exhibiting breath discoordination, characterized by irregular breathing patterns or breath-holding, were labeled separately (as “Uneven”), with further differentiation based on whether muscle activation remained symmetric (labelled as “Bal”, for “balanced”) or exhibited unilateral dominance. These distinctions resulted in six exercise categories, covering a range of breathing irregularities, asymmetric force application, or both.

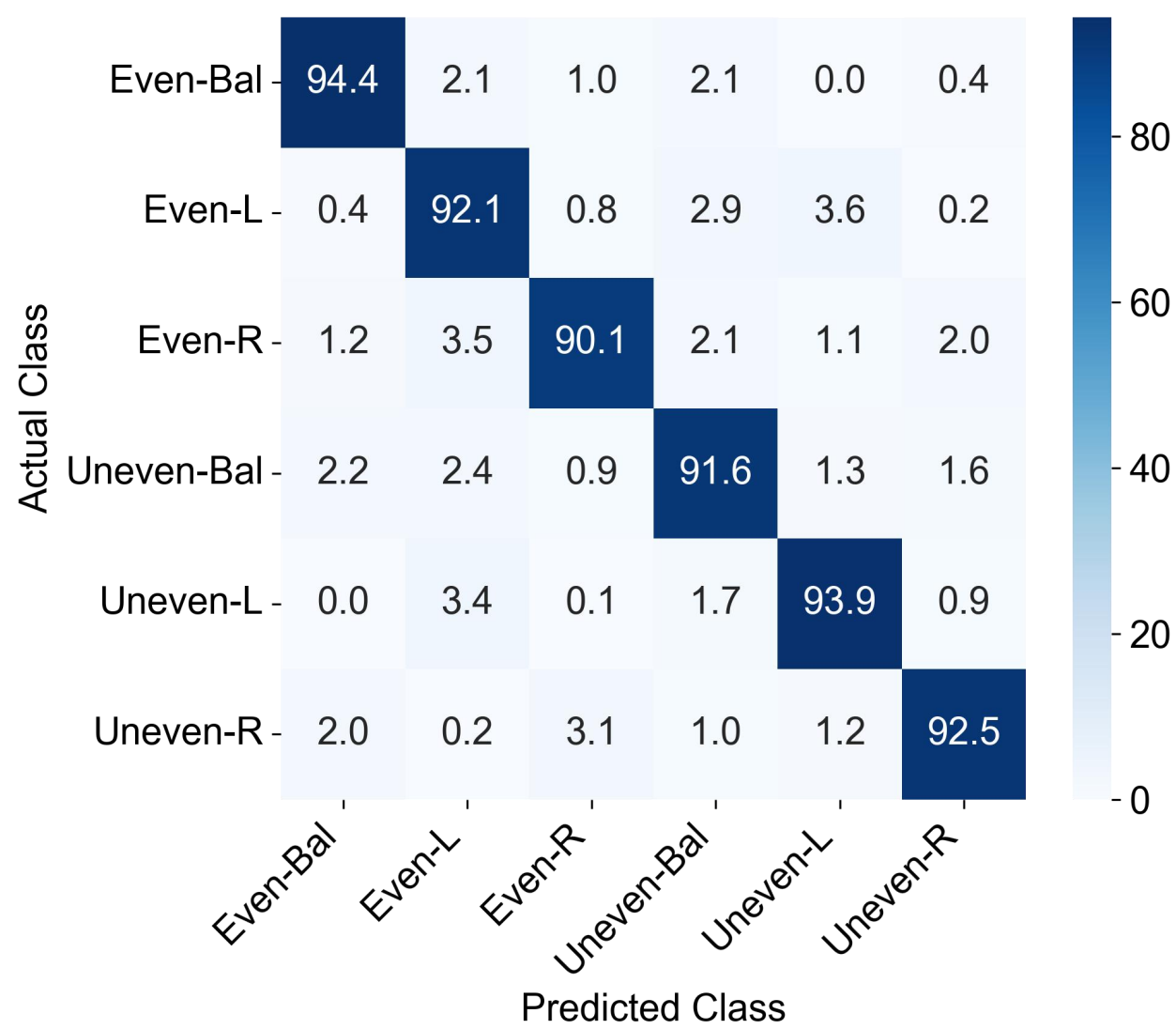

**Fig. 7. Confusion matrix of the classification model.**

This structured labeling approach provides a comprehensive foundation for training the classification model, allowing it to recognize subtle deviations from optimal lifting form. By systematically capturing improper breathing mechanics and force imbalances, the dataset enables the development of an intelligent feedback system capable of enhancing exercise execution and injury prevention. Training and validation accuracy/loss curves are shown in Figure 8, confirming convergence and absence of overfitting.

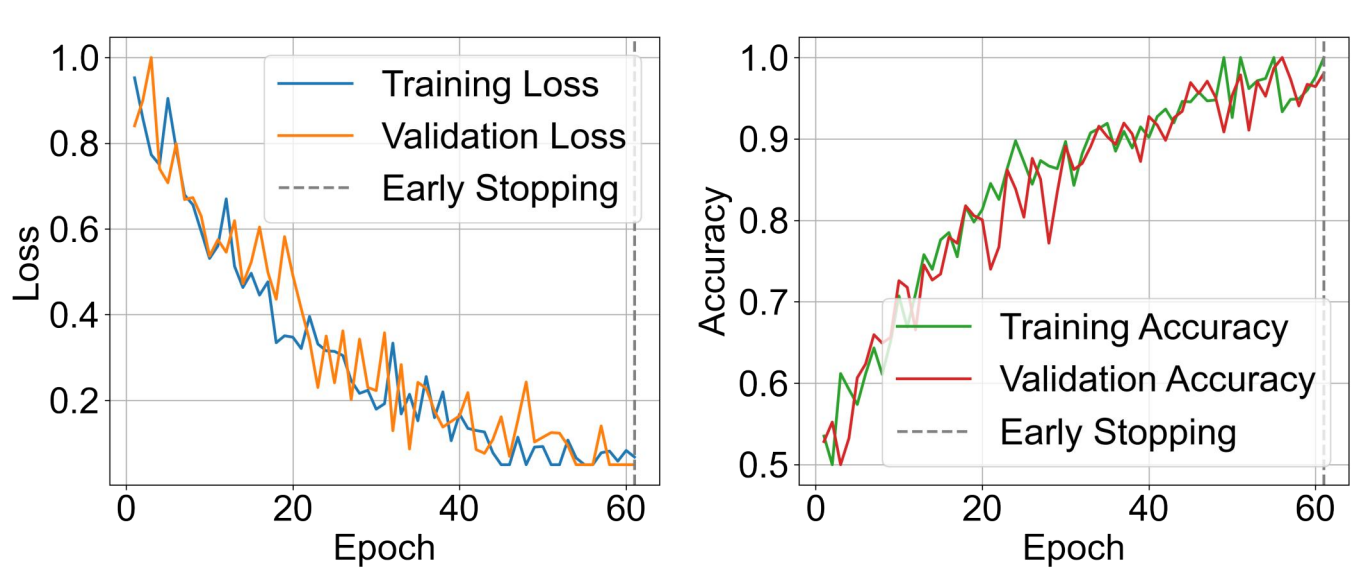

**Fig. 8. Training and validation loss and accuracy curves.**

*B. Model Performance*

To evaluate the classification accuracy of the proposed deep learning model, we computed the confusion matrix, as shown in Figure 7. The model demonstrated high classification accuracy across all six categories, with the majority of predictions aligning with the actual labels. The highest classification accuracy was achieved in the "Even-Bal" and "Uneven-L" classes, both reaching 94.2%, while the lowest accuracy was observed in the "Even-R" class, where 3.8% of instances were misclassified as "Even-L".

Misclassifications primarily occurred between adjacent categories, such as "Even-R" and "Even-L" or "Uneven-Bal" and "Uneven-R", indicating that the model occasionally struggled with borderline asymmetries in muscle activation. Despite this, the overall mean classification accuracy remained above 92%, demonstrating the model’s robustness in detecting breath-force coordination patterns and muscle activation asymmetry.

*C. Ablation Study*

To assess the effectiveness of different feature extraction models and the impact of preprocessing, we conducted an ablation study comparing multiple deep learning and machine learning classifiers. Figure 9 illustrates the classification accuracy of various models under two conditions: baseline (with preprocessing) and without filtering.

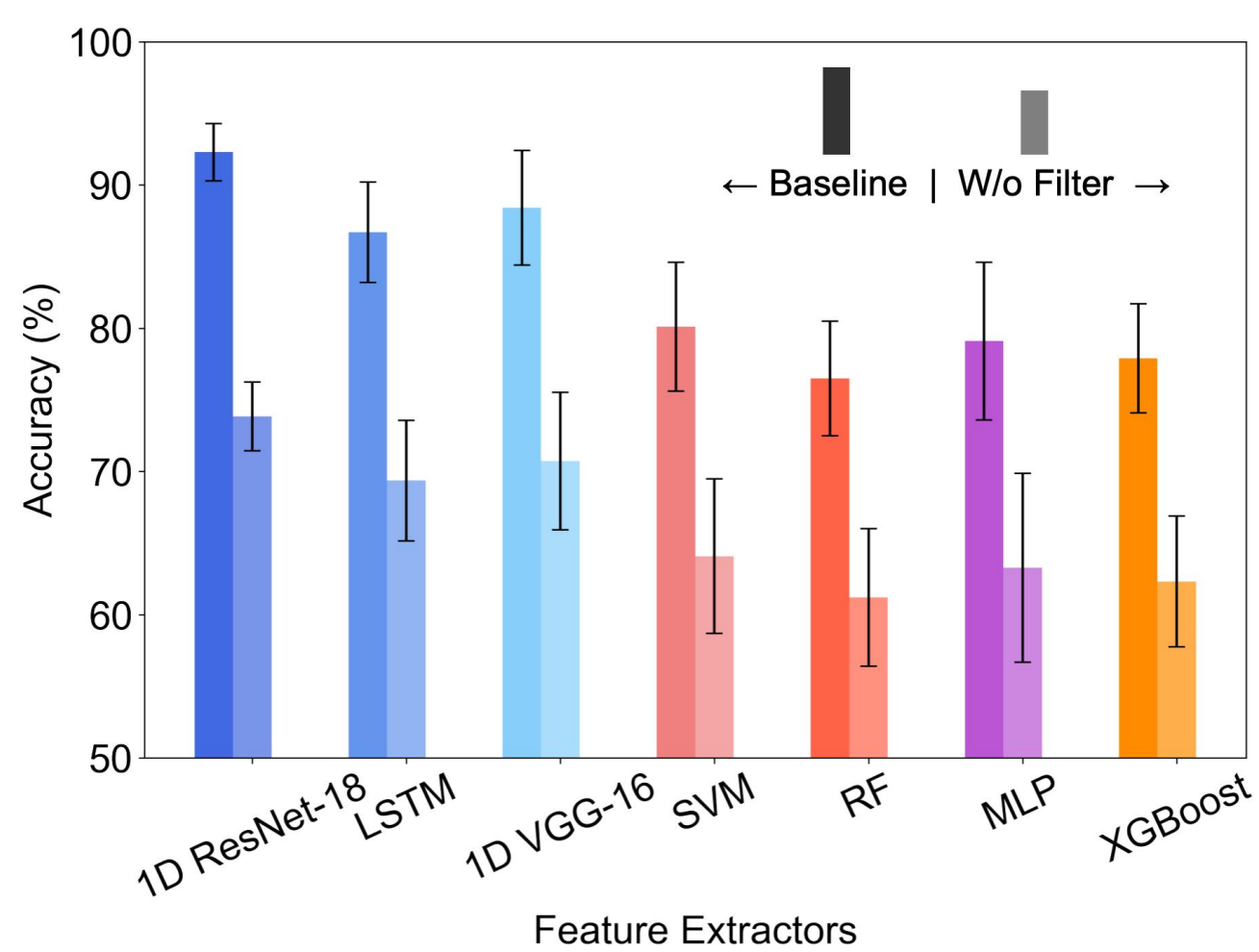

**Fig. 9. Ablation study comparing different feature extractors and evaluating the effect of filtering.**

Among the tested models, 1D ResNet-18 demonstrated the highest classification accuracy, achieving over 92% mean accuracy with preprocessing applied. This result indicates that deep residual networks effectively capture temporal dependencies in strain sensor signals, making it the optimal feature extraction backbone for our system. The LSTM and 1D VGG-16 models also performed well, but with slightly lower accuracy and higher variance, suggesting that although they can learn sequential dependencies, they are less efficient in handling biomechanical strain data compared to ResNet-based architectures.

Traditional machine learning models, including SVM, Random Forest (RF), and XGBoost, exhibited substantially lower accuracy, likely due to their inability to directly learn hierarchical temporal features from raw signals. Similarly, the MLP (Multi-Layer Perceptron) classifier underperformed compared to CNN-based models, highlighting the necessity of effective spatial-temporal feature extraction for breath-force and muscle activation classification.

Additionally, the comparison between baseline and non-filtered conditions reveals a clear accuracy drop across all models when low-pass filtering is removed. This decline suggests that high-frequency noise negatively impacts model performance, reinforcing the importance of our 10 Hz low-pass filter in preserving meaningful biomechanical signals. The significant accuracy reduction across all classifiers further confirms that the most informative signal components reside below 10 Hz, aligning with prior findings in human movement and respiratory signal analysis.

*D. Further Analysis of the Model*

To further investigate the model's feature representation, we applied t-distributed stochastic neighbor embedding (t-SNE) to visualize the separability of different exercise categories in the learned feature space. Figure 10 presents the t-SNE projection of raw signals (left) and extracted features (right) [37].

In the raw signal space, the six exercise categories exhibit significant overlap, indicating that raw strain signals alone do not provide clear decision boundaries for classification. After feature extraction using 1D ResNet-18, the separability of different categories improves considerably, demonstrating the effectiveness of the learned representations. However, some class clusters remain closer to one another, specifically (1,4), (2,5), and (3,6). This suggests that distinguishing between breath-force coordination errors (uneven vs. proper breathing) is more challenging than differentiating left-right asymmetries. This finding aligns with the confusion matrix results (Figure 7), where misclassifications are more frequent between categories that share similar muscle activation patterns but differ in breathing irregularities. This is likely due to the inherently higher inter-subject variability in breathing patterns compared to muscle activation, as well as the lower signal amplitude and more subtle temporal features associated with respiratory-induced deformations in the abdominal strain channel. Additionally, improper breathing may occur asynchronously or intermittently during repetitions, making it harder for the model to capture consistent patterns.

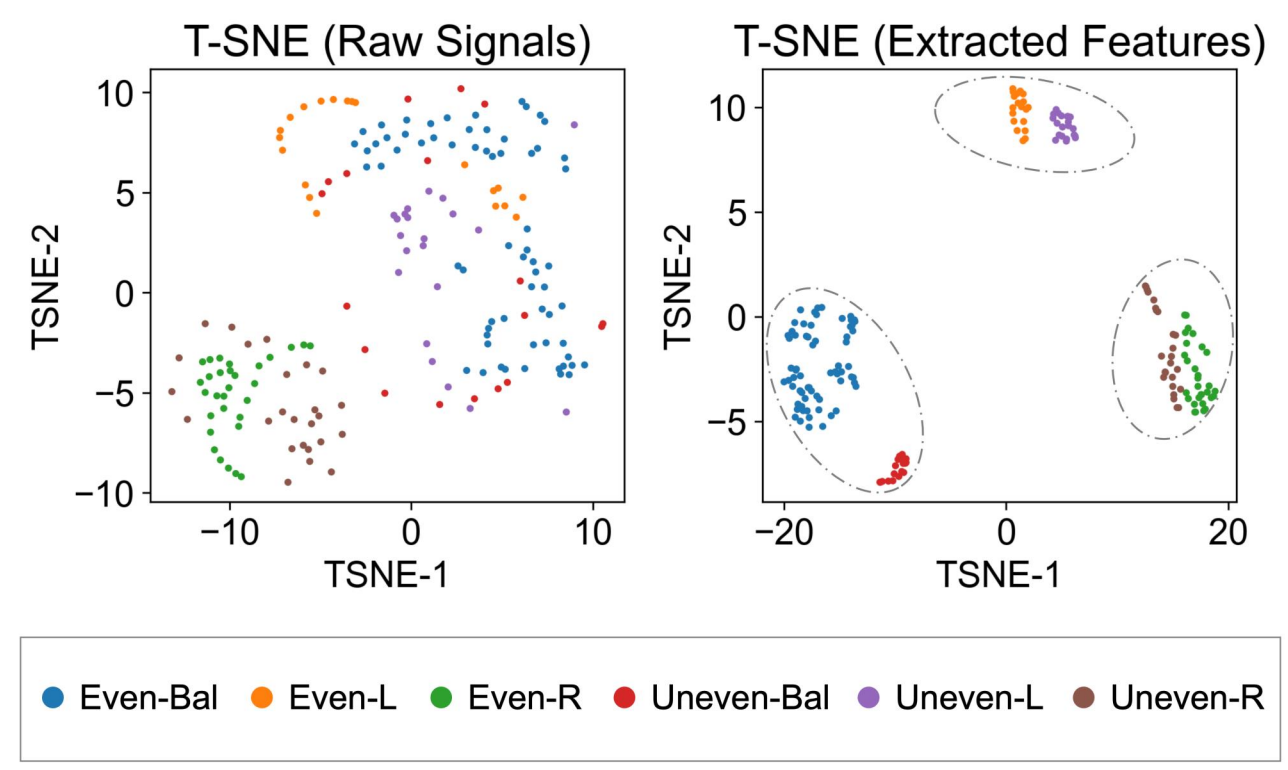

**Fig. 10. t-SNE visualization of six exercise categories.**

To further interpret the decision-making process of the deep learning model, we employed Grad-CAM (Gradient-weighted Class Activation Mapping) to visualize the network's attention when classifying different exercise forms [27]. Figure 11 illustrates how the model assigns importance to different time segments when distinguishing between Even Breathing with Balanced Muscle Activation and Uneven Breathing with Left-Dominant Muscle Activation. The background shading in the plots represents the network's attention distribution, where brighter (whiter) regions indicate areas of higher importance, and darker (grayer) regions correspond to areas with lower attention.

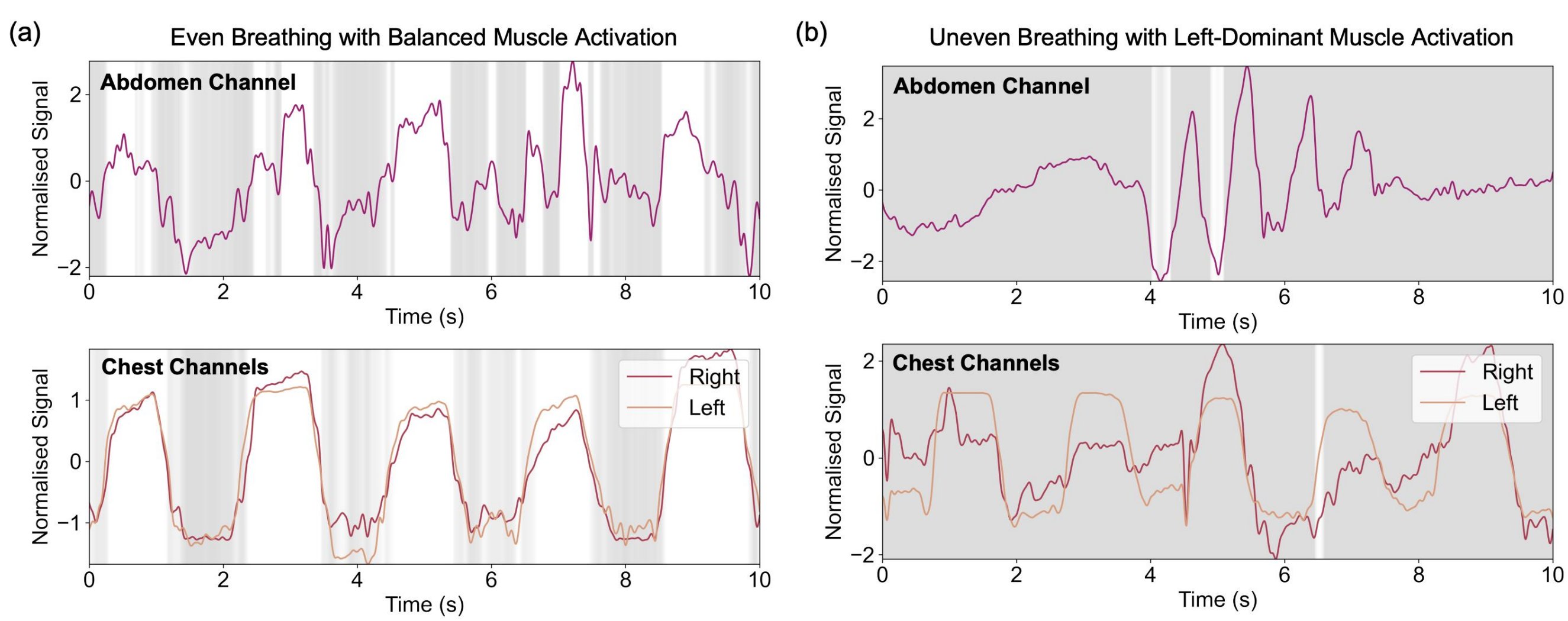

**Fig. 11. Grad-CAM-based explainability analysis of the classification model, highlighting attention regions for different exercise categories.** The background shading represents the network's attention distribution, where brighter (whiter) regions indicate higher attention. (a) A case of even breathing with balanced muscle activation, where the model consistently attends to periodic cycles to confirm proper execution. (b) A case of uneven breathing with left-dominant exertion, where the model selectively focuses on specific irregular segments to detect the asymmetry.

In Figure 11(a), which corresponds to a sample with even breathing and symmetric muscle exertion, the model distributes its attention consistently across periodic cycles, as it needs to capture the regularity in breathing and force symmetry to confirm proper form. This indicates that the model is correctly learning to focus on global rhythmic features when identifying well-coordinated movements.

In contrast, Figure 11(b) depicts a case of uneven breathing with excessive left-side exertion. Here, the model selectively focuses on specific irregular segments, particularly those with breathing inconsistencies and asymmetrical muscle activation. This suggests that for detecting improper execution, the model does not require a uniform attention distribution but rather emphasizes localized deviations from expected patterns, which aligns well with human intuition in recognizing movement inconsistencies.

The XAI visualization results confirm the strong interpretability of the network, showing that it accurately attends to biomechanically relevant features in both balanced and unbalanced movement cases. This reinforces confidence in the model's decision-making process and its potential for providing explainable, real-time feedback to users during strength training.

### *E. Real-world Deployment*

#### *1) Latency Analysis*

To substantiate the system's real-time performance, we quantified the end-to-end latency from strain signal acquisition to classification output. The latency measurement includes: (i) analog-to-digital signal acquisition, (ii) wireless data transmission, (iii) signal preprocessing, and (iv) model inference on a standard laptop (Intel i7-1165G7 CPU, 16 GB RAM).

Using oscilloscope-based trigger alignment and timestamp comparison on the receiving device, the measured average end-to-end latency was 68 ± 4 ms. The breakdown is as follows:

(1) Wi-Fi transmission and buffering: ~36 ms

(2) Signal preprocessing: ~10 ms

(3) Model inference on laptop: ~18 ms

(4) Other system overheads: ~4 ms

This latency is well below the 100 ms threshold commonly regarded as the upper bound for real-time biomechanical feedback, confirming that our system can deliver responsive, closed-loop guidance in practical fitness and rehabilitation scenarios.

#### *2) Power Consumption Evaluation*

To further assess real-world viability, we quantified the power consumption of the system under typical usage. The readout module, based on the ESP32 SoC, was configured to sample three analog channels at 130 Hz and stream data continuously via Wi-Fi. The system was powered by a 3.7V, 1200 mAh LiPo battery.

The measured power draw was:

(1) ESP32 sampling and ADC operation: ~83 mW

(2) Wi-Fi transmission (continuous streaming): ~102 mW

This corresponds to an estimated battery life of approximately 24 hours under continuous operation. Notably, all computation including signal preprocessing and neural network inference was performed on an external laptop and is not included in the wearable module's power budget. This setup ensures that the sensing garment remains lightweight and low-power, suitable for extended monitoring sessions without frequent recharging. Future iterations may incorporate low-power embedded inference hardware to further improve autonomy and system compactness.

#### *3) Displacement Issues*

To evaluate the system's robustness to common placement inconsistencies during garment donning, we conducted a sensor displacement study using three participants. Each subject performed bench press repetitions under three placement conditions: i) Baseline (ideal sensor alignment based on anatomical landmarks); ii) ± 10 mm horizontal displacement (simulating lateral shift of sensor positions due to incorrect shirt alignment); and iii) 10% rotational displacement (simulating angular misalignment around the body axis).

The recorded data from each condition were processed using the original trained model without retraining. Compared to the baseline, the ±10 mm horizontal shift resulted in an average classification accuracy drop of 6.1%, while rotational displacement led to a 4.8% reduction. Qualitatively, the abdominal channel (used for breath-force detection) exhibited moderate tolerance to both forms of displacement, whereas the pectoral channels (used for detecting muscle asymmetry) were more sensitive, especially under lateral misalignment. These results suggest that the system tolerates minor displacement to some extent, but consistent garment positioning remains critical for optimal classification performance. Future developments may explore automatic alignment calibration or adaptive preprocessing to enhance robustness.

### *F. Comparison with Established Modalities*

To evaluate the effectiveness of our textile strain sensor-based system against established sensor modalities, we conducted a head-to-head performance comparison with a combined IMU and EMG configuration. While previous studies have used IMU or EMG individually for exercise recognition, these approaches typically focus on either joint kinematics or localized muscle activity, and are not capable of jointly capturing respiratory-muscular coordination or bilateral muscle activation asymmetry. Our system is designed to integrate these two types of information using a unified strain-sensing mechanism embedded in standard sportswear, offering both physiological relevance and practical wearability.

To create a baseline for comparison, we collected data from three participants (subjects ID: 1, 2, 3) performing the same six exercise types used in the main study. Alongside our strain-sensing garment, participants were equipped with a commercial EMG module (SeeedStudio) and the OpenBCI EmotiBit IMU. The EMG electrodes were placed symmetrically over the left and right pectoralis major muscles to record activation asymmetry, and the IMU was mounted on the upper abdomen to capture torso dynamics related to respiratory motion. All signals were processed through the same classification pipeline and fed into the same 1D ResNet-18 model architecture to ensure comparability.

The classification accuracy achieved using our textile strain sensor system was 94.1% ± 1.7%. In contrast, the IMU+EMG configuration reached 92.8% ± 2.1% accuracy under identical training and evaluation settings. While the difference is modest, it demonstrates that the textile-based method can match or slightly outperform conventional multi-sensor setups, without the added burden of skin-contact electrodes or body-mounted rigid devices. We observed that the EMG signal was highly sensitive to skin-electrode impedance variability and susceptible to motion artifacts, while the IMU signal suffered from inter-subject inconsistency in respiratory dynamics. In contrast, our textile strain sensors provided more consistent signal quality across users, likely contributing to the improved classification results.

These findings validate the sensing capacity of our wearable system and reinforce its potential as a viable alternative to traditional wearable sensing technologies. The comparison also highlights the trade-off between sensor modality complexity and end-user comfort in practical deployment scenarios.

## IV. Discussion

This study presents a wearable smart sportswear system that integrates ultrasensitive textile strain sensors with wireless signal acquisition and deep learning-based classification to monitor breath-force coordination and muscle activation asymmetry during strength training. Through screen-printed graphene sensors strategically placed on the pectoralis major and upper abdomen, the system effectively captures respiratory and muscle activity patterns, enabling real-time biomechanical assessment. Our results demonstrate that the 1D ResNet-18 model serves as an optimal feature extraction backbone, achieving an average classification accuracy of 92.1% across six exercise categories. Further t-SNE analysis confirmed that while muscle asymmetries are well-distinguished, breath-force irregularities remain more challenging to classify, aligning with confusion matrix findings. Additionally, Grad-CAM-based XAI visualization validated the interpretability of the network, ensuring that its attention aligns with biomechanically relevant features, reinforcing its applicability for real-time exercise guidance and injury prevention.

Despite these promising results, several limitations and broader application opportunities remain. First, while the wireless signal acquisition module successfully enables real-time monitoring, the current implementation relies on a relatively bulky PCB-based system. Future iterations may adopt flexible printed circuits (FPCs) or fully textile-integrated electronics to improve wearability and user comfort without compromising signal fidelity [38, 39].

Beyond gym-based training, the system's untethered architecture and fabric-integrated sensors inherently support deployment in unstructured, free-movement environments. Future studies will explore its use during unsupervised rehabilitation or natural daily activities by incorporating diverse movement data and retraining the model for generalization beyond predefined templates.

Additionally, individuals undergoing rehabilitation, such as those recovering from stroke or musculoskeletal injuries, represent a critical application population. Although this study is limited to healthy participants due to the absence of clinical ethical approval, future work will focus on obtaining such approvals and validating the system in clinical rehabilitation settings. The ability to non-invasively monitor asymmetric muscle recruitment and irregular breathing patterns could provide actionable insights for therapists, enabling dynamic feedback during physiotherapy or home-based recovery. In parallel, the same sensing architecture may also benefit long-term postural and respiratory monitoring in sedentary or elderly populations.

Lastly, integrating large language model (LLM) agents and human body digital twin (DT) technologies could further refine real-time exercise assessment. These capabilities may enable adaptive feedback systems that dynamically adjust recommendations based on individual biomechanical patterns and longitudinal training data [5, 40, 41, 42]. This integration could support the development of comprehensive digital profiles, bridging the gap between AI-driven motion analysis and personalized athletic coaching.

By addressing these challenges and extending the system beyond constrained lab conditions, future developments will enable the seamless fusion of wearable biosensing, real-time AI-driven feedback, and digital human modeling, paving the way for next-generation intelligent sportswear that supports personalized, data-driven training, rehabilitation, and health optimization across diverse real-world settings.

## V. Conclusion

This work presents a wearable smart sportswear system that integrates ultrasensitive textile strain sensors with a wireless deep learning-based framework for real-time monitoring of breath-force coordination and muscle activation symmetry. By screen-printing graphene-based strain sensors onto compression sportswear and the 1D ResNet-18 architecture for feature extraction, the system achieves 92.1% classification accuracy, ensuring high-fidelity biomechanical assessment during strength training. Its wireless and real-time AI-driven architecture enhances usability in exercise optimization, rehabilitation, and athletic performance monitoring, paving the way for next-generation intelligent sportswear with the potential for integration into digital twin modeling, adaptive feedback systems, and AI-powered human movement analysis.